\documentclass[twocolumn,showpacs,preprintnumbers,amsmath,amssymb]{revtex4-1}

\usepackage{afterpage}
\usepackage[dvipdfmx]{graphicx}
\usepackage{color,bm}

\begin{document}

\title{Collinear and triangular solutions to the 
coplanar and circular three-body problem in the parametrized post-Newtonian formalism}
\author{Yuya Nakamura}
\email{nakamura@tap.st.hirosaki-u.ac.jp}
\author{Hideki Asada} 
\email{asada@hirosaki-u.ac.jp}
\affiliation{
Graduate School of Science and Technology, 
Hirosaki University,
Hirosaki 036-8561, Japan} 

\date{\today}

\begin{abstract}
This paper investigates the 
coplanar and circular 
three-body problem in the parametrized post-Newtonian (PPN) formalism, 
for which we focus on a class of fully conservative theories 
characterized by the Eddington-Robertson parameters $\beta$ and $\gamma$. 
It is shown that there can still exist a collinear equilibrium configuration and a triangular one, 
each of which is a generalization of the post-Newtonian equilibrium configuration in general relativity. 
The collinear configuration 
can exist for arbitrary mass ratio, $\beta$, and $\gamma$. 
On the other hand, 
the PPN triangular configuration depends on the nonlinearity parameter $\beta$ 
but not on $\gamma$. 
For any value of $\beta$, 
the equilateral configuration is possible,  
if and only if 
three finite masses are equal or two test masses orbit around one finite mass. 
For general mass cases, 
the PPN triangle is not equilateral 
as in the post-Newtonian case. 
It is shown also that 
the PPN displacements from the Lagrange points 
in the Newtonian gravity 
$L_1$, $L_2$ and $L_3$ 
depend on $\beta$ and $\gamma$, 
whereas those to $L_4$ and $L_5$ 
rely only on $\beta$. 
\end{abstract}


\maketitle

\section{Introduction}
The three-body problem is among the classical ones in physics.  
It led to the notion of the chaos 
\cite{Goldstein}. 
On the other hand, particular solutions 
such as 
Euler's collinear solution and Lagrange's equilateral one 
\cite{Danby,Marchal} 
express regular orbits and they have still attracted interest 
e.g. \cite{Asada,Torigoe,Seto,Schnittman,Connors}. 
If one mass is zero and the other two masses are finite,  
the collinear solution and triangular one 
correspond to Lagrange points 
$L_1$, $L_2$, $L_3$, $L_4$ and $L_5$ 
as particular solutions 
for the coplanar restricted three-body problem. 

In his pioneering work \cite{Nordtvedt}, 
Nordtvedt found that 
the position of the triangular points 
is very sensitive to the ratio of the gravitational mass 
to the inertial mass in gravitational experimental tests, 
where the post-Newtonian (PN) terms are not fully taken into account. 

Krefetz \cite{Krefetz} and Maindl \cite{Maindl} 
studied 
the restricted three-body problem in the PN approximation 
and found the PN triangular configuration for 
a general mass ratio between two masses. 
These investigations were extended to 
the PN three-body problem for 
general masses 
\cite{Yamada2010,Yamada2011,Ichita2011,Yamada2012,Yamada2015,Yamada2016}, 
and the PN counterparts for 
Euler's collinear \cite{Yamada2010,Yamada2011} 
and Lagrange's equilateral solutions \cite{Ichita2011,Yamada2012}
were obtained. 
It should be noted that 
the PN triangular solutions are not necessarily equilateral 
for general mass ratios 
and they are equilateral only for either the equal mass case 
or two test masses. 
The stability of the PN solution  and 
the radiation reaction at 2.5PN order 
were also examined \cite{Yamada2015,Yamada2016}. 

In a scalar-tensor theory of gravity, 
a collinear configuration for three-body problem 
was discussed 
\cite{Zhou}. 
In addition to such fully classical treatments, 
a possible quantum gravity correction to the Lagrange points was proposed  \cite{Battista2015a,Battista2015b}. 

Moreover, the recent discovery of a relativistic hierarchical triple system 
including a neutron star \cite{Ransom} 
has generated  
renewed interest in the relativistic three-body problem 
and the related gravitational experiments  
\cite{Archibald,Will2018,Voisin}. 

The main purpose of the present paper is 
to reexamine the coplanar and circular three-body problem 
especially in the PPN formalism. 
One may ask if 
collinear and triangular configurations 
are still solutions 
for the coplanar three-body problem in the PPN gravity. 
If so, 
how large are the PPN effects of the three-body configuration? 
We focus on the Eddington-Robertson parameters $\beta$ and $\gamma$, 
because the two parameters are the most important ones; 
$\beta$ measures how much nonlinearity there is in the superposition law for gravity 
and $\gamma$ measures how much space curvature is produced by unit rest mass 
\cite{Will,Poisson} . 
Hence, preferred locations, preferred frames or 
a violation of conservation of total momentum will not be considered 
in this paper. 
We confine ourselves to a class of fully conservative theories. 
See e.g. \cite{Klioner} for the celestial mechanics in this class of PPN theories. 

This paper is organized as follows. 
In Section II, 
collinear configurations are discussed in the PPN formalism. 
Section III investigates 
PPN triangular configurations. 
In Section IV, 
the PPN corrections to the Lagrange points  
are examined. 
For brevity, the Lagrange points defined in Newtonian gravity 
are referred to as the Newtonian Lagrange points in this paper. 
Section V summarizes this paper. 
Throughout this paper, $G=c=1$. 
$A, B$ and $C \in \{1,2,3\}$ label three masses.

\section{Collinear configuration in PPN gravity}
\subsection{Euler's collinear solution in Newton gravity}
Let us begin with briefly mentioning the Euler's collinear solution 
for the circular three-body problem in Newton gravity 
\cite{Danby,Marchal}, 
for which each mass $M_A$ ($A = 1, 2, 3$) at $\bm{x}_A$ is orbiting 
around the common center of mass (COM) at $\bm{x}_G$, 
and the orbital velocity and acceleration are denoted 
as $\bm{v}_A$ and $\bm{a}_A$, respectively.
In this section, 
we suppose that three masses are always aligned, 
for which it is convenient to use the corotating frame 
with a constant angular velocity $\omega$ 
on the orbital plane chosen as the $x-y$ plane. 

\begin{figure}
\includegraphics[width=7.5cm]{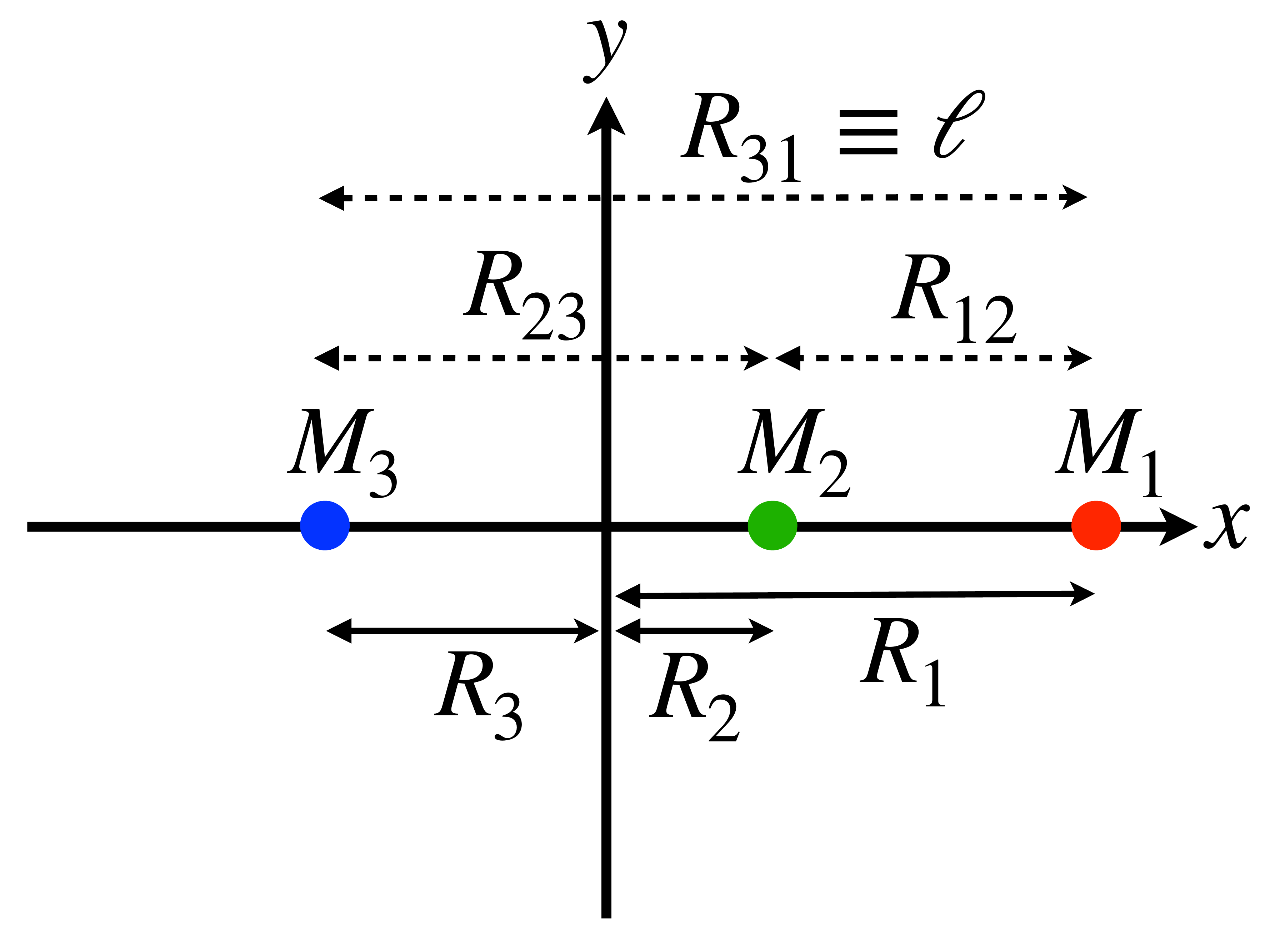}
\caption{
Schematic figure for the collinear configuration of three masses. 
}
\label{figure-collinear}
\end{figure}

Without loss of generality, we assume $x_1 > x_2 > x_3$ 
for $\bm{x}_A \equiv (x_A, 0)$. 
Let $R_A$ denote the relative position of each mass $M_A$ from the COM 
at $\bm{x}_G \equiv (x_G, 0)$. 
Namely, $R_A = x_A -x_G$. 
Note that $|R_A| \neq |\bm{x}_A|$ unless $x_G = 0$ is chosen.  
We define the relative vector between masses as 
$\bm{R}_{AB} \equiv \bm{x}_A - \bm{x}_B$, 
for which the relative length is $R_{AB} = |\bm{R}_{AB}|$.  
See Figure \ref{figure-collinear} for a configuration of the Euler's collinear solution.

The coordinate origin $x=0$ is chosen between $M_1$ and $M_3$, 
such that $R_1 > R_2 > R_3$, $R_1 > 0$ and $R_3 <0$. 
By taking account of this sign convention, 
the equation of motion becomes 
\begin{eqnarray}
R_1 \omega^2 &=& \frac{M_2}{R_{12}^2} + \frac{M_3}{R_{13}^2} , 
\label{EOM-M1-N}
\\
R_2 \omega^2 &=& -\frac{M_1}{R_{12}^2} + \frac{M_3}{R_{23}^2} , 
\label{EOM-M2-N}
\\
R_3 \omega^2 &=& -\frac{M_1}{R_{13}^2} - \frac{M_2}{R_{23}^2} . 
\label{EOM-M3-N} 
\end{eqnarray}

We define the distance ratio as $z \equiv R_{23}/R_{12}$, 
which plays a key role in the following calculations. 
Note that $z > 0$ by definition. 
We subtract Eq. ($\ref{EOM-M2-N}$) from Eq. ($\ref{EOM-M1-N}$) 
and Eq. ($\ref{EOM-M3-N}$) from Eq. ($\ref{EOM-M2-N}$). 
By combining the results including the same angular velocity $\omega$, 
we obtain a fifth-order equation for $z$ as
\begin{align}
&(M_1+M_2) z^5 + (3M_1+2M_2)z^4 + (3M_1+M_2)z^3 
\notag\\
&- (M_2+3M_3)z^2 - (2M_2+3M_3)z - (M_2+M_3) = 0 , 
\label{5th}
\end{align}
for which there exists the only positive root \cite{Danby,Marchal}. 
In order to obtain Eq. (\ref{5th}), 
we do not have to specify the coordinate origin e.g. $x_G = 0$. 
This is because Eq. (\ref{5th}) does not refer to any coordinate system. 
Once Eq. (\ref{5th}) is solved for $z$, 
we can obtain $\omega$ by substituting $z$ into any of 
Eqs. (\ref{EOM-M1-N})-(\ref{EOM-M3-N}).

%%%%%
\subsection{PPN collinear configuration}
In a class of fully conservative theories 
including only the Eddington-Robertson parameters $\beta$ and $\gamma$,  
the equation of motion is 
\cite{Will,Poisson} 
\begin{align}
\bm{a}_A =& - \sum_{B\neq A} \frac{M_B}{R_{AB}^2} \bm{n}_{AB} 
          \notag\\
          & - \sum_{B\neq A } \frac{M_B}{R_{AB}^2} \bigg\{ \gamma v_A^2 - 2(\gamma + 1)(\bm{v}_A \cdot \bm{v}_B) \nonumber\\
        &~~~~~ + (\gamma + 1)v_B^2 
         - \frac{3}{2} (\bm{n}_{AB} \cdot \bm{v}_B)^2      
         - \bigg(2\gamma + 2\beta +1  \bigg) \frac{M_A}{R_{AB}} 
         \notag\\
         &~~~~~
       - 2(\gamma +\beta) \frac{M_B}{R_{AB}} 
         \bigg\} \bm{n}_{AB} \nonumber\\  
       & + \sum_{B\neq A} \frac{M_B}{R_{AB}^2} \bigg\{\bm{n}_{AB} \cdot [2(\gamma +1) \bm{v}_A - (2\gamma+1) \bm{v}_B]
         \bigg\}(\bm{v}_A - \bm{v}_B) \nonumber\\       
      & + \sum_{B \neq A} \sum_{C\neq A, B} \frac{M_B M_C}{R_{AB}^2} 
      \bigg[ \frac{2(\gamma +\beta)}{R_{AC}} +  \frac{2\beta -1}{R_{BC}} 
      \notag\\
      &~~~~~~~~~~~~~~~~~~~~~~~~~~~~~~ 
      -\frac{1}{2}  \frac{R_{AB}}{R_{BC}^2} (\bm{n}_{AB}\cdot\bm{n}_{BC}) 
      \bigg] \bm{n}_{AB} \nonumber\\
    &- \frac{1}{2} (4\gamma +3 ) 
     \sum_{B\neq A} \sum_{C\neq A,B} \frac{M_B M_C}{R_{AB} R_{BC}^2} \bm{n}_{BC}
          + O(c^{-4}) , 
         \label{PPN-EOM}
\end{align}
where 
\begin{align}
\bm{n}_{AB} &\equiv \frac{\bm{R}_{AB}}{R_{AB}} . 
\label{n}
\end{align}

For three aligned masses,   
Eq. (\ref{PPN-EOM})  becomes 
the force-balance equation as 
\begin{align}
\ell \omega^2 = F_N + F_M + F_V \omega^2 ,
\label{EOM-PPN-collinear}
\end{align}
where 
we define 
$\ell \equiv R_{31}$, 
the mass ratio $\nu_A \equiv M_A/M$ for $M \equiv \sum_A M_A$, 
and 
\begin{align}
F_N = \frac{M}{\ell^2 z^2} [
            &1 - \nu_1 - \nu_3 + 2  (1 - \nu_1 - \nu_3) z 
            + (2 - \nu_1 - \nu_3) z^2  \nonumber\\
            &+ 2 (1 - \nu_1 - \nu_3) z^3
            + (1 - \nu_1 - \nu_3) z^4 ] ,
\end{align}
\begin{align}
F_M = - \frac{M^2}{\ell^3 z^3} [
           &\{2 (\beta + \gamma) \nu_2 + (1 + 2\beta + 2\gamma) \nu_3 \} \nu_2 \nonumber\\
           &+ \{(-1 + 4\beta + 2\gamma) \nu_1 + 6 (\beta + \gamma) \nu_2 
           \notag\\
           &~~~~~ + 3 (1 + 2\beta + 2\gamma) \nu_3 \} \nu_2 z 
           \nonumber\\
           &+ \{(-5 + 12 \beta +4 \gamma) \nu_1 + 6 (\beta +\gamma) \nu_2 
           \notag\\
           &~~~~~ - (1 - 10 \beta - 4 \gamma) \nu_3 \} \nu_2 z^2 
           \nonumber\\
           &+ \{2 (\beta + \gamma) \nu_1^2 + 4 (\beta + \gamma) \nu_2^2 
           \notag\\
           &~~~~~ - (7 - 14 \beta - 2 \gamma) \nu_2 \nu_3 
           + 2 (\beta + \gamma) \nu_3^2
           \nonumber\\
           & + ((-7 + 14 \beta + 2 \gamma) \nu_2 
           \notag\notag\\
           &~~~~~  + 2 (1 + 2\beta + 2 \gamma) \nu_3) \nu_1 \} z^3 \nonumber\\
           & + \{(-1 + 10 \beta + 4 \gamma) \nu_1 + 6 (\beta + \gamma) \nu_2 
           &\notag\\
           &~~~~~ + (12 \beta + 4\gamma -5) \nu_3 \} \nu_2 z^4 
           \nonumber\\
           &+ \{ 3( 1 + 2 \beta + 2\gamma) \nu_1 + 6 (\beta + \gamma) \nu_2 
           \notag\\
           &~~~~~  + (-1 +4 \beta + 2\gamma) \nu_3 \} \nu_2 z^5 
           \nonumber\\
           &+ \{(1 + 2\beta + 2 \gamma) \nu_1 + 2 (\beta + \gamma) \nu_2 \} \nu_2 z^6
           ] ,
\end{align}
and 
\begin{align} 
F_V =& \frac{M}{(1+z)^2 z^2}
\notag\\
&\times[
           - \nu_1^2 \nu_2 
           -2 \nu_1 \nu_2 (2 \nu_1 + \nu_2) z \nonumber\\
           &~~~ + \{\gamma \nu_1^3 +  ((-2 + 4 \gamma) \nu_2 + 3 (1 + \gamma) \nu_3 ) \nu_1^2 \nonumber\\
           &~~~ + (2 \nu_2 + \nu_3) (\gamma \nu_2^2 + (1 + 2 \gamma) \nu_2 \nu_3 + \gamma \nu_3^2) \nonumber\\
           &~~~ + ((-1 + 5 \gamma) \nu_2^2 + 8 (1 + \gamma) \nu_2 \nu_3 + 3 (1 + \gamma) \nu_3^2 ) \nu_1 \} z^2 
           \nonumber\\
           &~~~ + 2 (\nu_1 + 2 \nu_2 + \nu_3) \{
           \gamma \nu_1^2 + \gamma \nu_2^2 + (1 + 2\gamma) \nu_2 \nu_3  \nonumber\\
           &~~~ + \gamma \nu_3^2+ ((1 + 2 \gamma) \nu_2 + (3 + 2 \gamma) \nu_3) \nu_1 \} z^3 \nonumber\\
           &~~~ +\{ \gamma \nu_1^3 + 2 \gamma \nu_2^3 - (1 - 5\gamma) \nu_2^2 \nu_3 
           -2 (1 - 2 \gamma) \nu_2 \nu_3^2  \nonumber\\
           &~~~ + \gamma \nu_3^3+ ((1 + 4 \gamma) \nu_2 + 3 (1 + \gamma) \nu_3) \nu_1^2  \nonumber\\
           &~~~ + ((2 + 5 \gamma) \nu_2^2 + 8 (1 + \gamma) \nu_2 \nu_3 + 3 (1 + \gamma) \nu_3^2 ) \nu_1
           \} z^4 \nonumber\\
           &~~~ - 2 \nu_2 \nu_3 (\nu_2 + 2 \nu_3) z^5 
           - \nu_2 \nu_3^2 z^6
           ] .
\end{align}

By rearranging Eq. (\ref{PPN-EOM}) 
for the collinear configuration 
by the same way as in subsection II.A, 
we find a seventh-order equation for $z$ as 
\begin{align}
\sum_{k=0}^{7} A_k z^k = 0 ,
\label{7theq}
\end{align}
where the coefficients are 
\begin{align}
A_7 =& \frac{M}{\ell} \bigg[ 
-2(\beta + \gamma) - 2 \nu_1 +4 (\beta + \gamma) \nu_3  
+ 2 \nu_1^2 
+  4 \nu_1 \nu_3 
\notag\\
&~~~~~ -2 (\beta + \gamma) \nu_3^2 
-  2 \nu_1^2 \nu_3 
-  2 \nu_1 \nu_3^2
 \bigg] , 
\label{A7}
\end{align}
\begin{align}
A_6 = & 1 - \nu_3 
\notag\\
&+ \frac{M}{\ell} \bigg[ 
- (6\beta + 7\gamma) - (6 + 2\beta +2\gamma )\nu_1 
\notag\\
&~~~~~~~~ - (2-8\beta-11\gamma) \nu_3 
+ 4 \nu_1^2 
+ (12+2\beta+2\gamma)\nu_1\nu_3 
\notag\\
&~~~~~~~~ + (4-2\beta -4\gamma ) \nu_3^2 
+ 2 \nu_1^3 -  4 \nu_1^2 \nu_3 
-  6 \nu_1\nu_3^2
- 2 \nu_3^3
\bigg] , 
\label{A6}
\end{align}
\begin{align}
A_5 = & 2 + \nu_1 -2\nu_3 
\notag\\
&+\frac{M}{\ell} \bigg[
-3(2\beta +3\gamma) - 3 (2+2\beta +2\gamma ) \nu_1 
 - (6 - 11\gamma )\nu_3 
 \notag\\
&~~~~~~~~  +(12 +6\beta +2\gamma ) \nu_1\nu_3
+ (12 +6\beta -2\gamma ) \nu_3^2 
\notag\\
&~~~~~~~~  
+ 6 \nu_1^3
- 6 \nu_1 \nu_3^2 
- 6 \nu_3^3
\bigg] , 
\label{A5}
\end{align}
\begin{align}
A_4 = & 1 + 2\nu_1 - \nu_3 
\notag\\
&
+ \frac{M}{\ell} \bigg[ 
-2\beta -4 \gamma - (2\beta +8\gamma ) \nu_1 
- (6 +6\beta -8\gamma )\nu_3 
\notag\\
&~~~~~~~~  
 - (6 + 4\beta -2\gamma  ) \nu_1^2 
+ (4 +2\beta -2\gamma ) \nu_1 \nu_3 
  \notag\\
&~~~~~~~~
+ (12 +8\beta -4\gamma ) \nu_3^2 
+  6 \nu_1^3 + 2 \nu_1^2 \nu_3 
- 4 \nu_1 \nu_3^2
- 6 \nu_3^3
\bigg] ,
\label{A4}
\end{align}
\begin{align}
A_3 =&  -1 +\nu_1 - 2\nu_3 
 \notag\\
&
+ \frac{M}{\ell} \bigg[
2\beta + 4\gamma + (6 + 6\beta - 8\gamma ) \nu_1 
+ (2\beta + 8\gamma ) \nu_3 
\notag\\
&~~~~~~~~
- (12 + 8\beta -4 \gamma ) \nu_1^2 
- (4 +2\beta -2\gamma )\nu_1 \nu_3 
 \notag\\
&~~~~~~~~
+ (6 +4\beta -2\gamma ) \nu_3^2 
+ 6 \nu_1^3 + 4 \nu_1^2 \nu_3 
- 2 \nu_1 \nu_3^2 
- 6 \nu_3^3
\bigg] ,
\label{A3}
\end{align}
\begin{align}
A_2 =  &-2 +2\nu_1 - \nu_3 
 \notag\\
&
+ \frac{M}{\ell} \bigg[
6\beta +9\gamma + (6 - 11\gamma ) \nu_1
+ (6 + 6\beta +6\gamma ) \nu_3  
 \notag\\
&~~~~~~~~
- (12 + 6\beta -2 \gamma ) \nu_1^2 
 - (12 + 6\beta +2\gamma ) \nu_1 \nu_3
 \notag\\
&~~~~~~~~
+ 6 \nu_1^3
+ 6 \nu_1^2 \nu_3 
- 6 \nu_3^3
\bigg] ,
\label{A2}
\end{align}
\begin{align}
A_1 = & -1 +\nu_1 
 \notag\\
&
+ \frac{M}{\ell} \bigg[
6\beta + 7 \gamma + (2 - 8 \beta -11 \gamma ) \nu_1 
 + (6 + 2\beta +2\gamma ) \nu_3 
  \notag\\
&~~~~~~~~
  - (4 -2 \beta -4 \gamma ) \nu_1^2
 - (12 +2 \beta + 2\gamma ) \nu_1 \nu_3
 \notag\\
&~~~~~~~~
 - 4  \nu_3^2  
+ 2 \nu_1^3 
+ 4  \nu_1 \nu_3^2 
 + 6  \nu_1^2 \nu_3
-  2  \nu_3^3
 \bigg] , 
 \label{A1}
\end{align}
\begin{align}
A_0 = & \frac{M}{\ell} \bigg[
2\beta +2\gamma - 4(\beta +\gamma) \nu_1 + 2  \nu_3 + 2(\beta +\gamma ) \nu_1^2
 \notag\\
&~~~~~~
 - 4 \nu_1 \nu_3 - 2  \nu_3^2 
 + 2  \nu_1^2 \nu_3 + 2  \nu_1 \nu_3^2
\bigg] .
\label{A0}
\end{align}
It follows that Eq. (\ref{7theq}) recovers 
 the PN collinear configuration by  
 Eq. (13) of Reference \cite{Yamada2011} 
if and only if $\beta = \gamma = 1$. 
The uniqueness is because the number of the parameters 
$\beta$, $\gamma$ is two 
for eight coefficients $A_0, \cdots, A_7$. 

From Eq. (\ref{EOM-PPN-collinear}) for $z$ obtained above, 
the angular velocity $\omega_{PPN}$ 
of the PPN collinear configuration is obtained as 
\begin{align}
\omega_{PPN} = \omega_{ N} \bigg( 
                 1 + \frac{F_M}{2 F_N}  + \frac{F_V}{ 2 \ell}\bigg) , 
                 \label{omega-collinear}
\end{align}
where $\omega_N = (F_N / \ell) ^{1/2}$ 
is the Newtonian angular velocity. 
The subscript $N$ denotes the Newtonian case.

%%%%%
\section{Triangular configuration in PPN gravity}
\subsection{Lagrange's equilateral solution in Newtonian gravity}
In this subsection, 
we suppose that the three masses are in 
coplanar and circular motion 
with keeping the same separation between the masses, 
namely $R_{AB} = a$ for a constant $a$.  

It is convenient to choose the coordinate origin as the COM, 
\begin{equation}
\sum_A M_A \bm{x}_A = 0 , 
\end{equation}
for which 
the equation of motion for each mass in the equilateral triangle configuration 
takes a compact form as  
\cite{Danby} 
\begin{equation}
\frac{d^2 \bm{x}_A}{dt^2} 
= - \frac{M}{a^3} \bm{x}_A . 
\label{N-EOM}
\end{equation}
See e.g. Eq. (8.6.5) in Reference \cite{Danby} for the derivation of Eq. (\ref{N-EOM}). 
A triangular configuration is a solution, 
if the Newtonian angular velocity $\omega_N$ satisfies 
\begin{equation}
(\omega_N)^2 = \frac{M}{a^3} . 
\label{omega2}
\end{equation}

The orbital radius $\ell_A$ of each mass 
around the COM is \cite{Danby} 
\begin{align}
\ell_1 &= a \sqrt{\nu_2^2 + \nu_2\nu_3 + \nu_3^2} , 
\label{N-r1}
\\
\ell_2 &= a \sqrt{\nu_1^2 + \nu_1\nu_3 + \nu_3^2} ,
\label{N-r2}
\\
\ell_3 &= a \sqrt{\nu_1^2 + \nu_1\nu_2 + \nu_2^2} . 
\label{N-r3}
\end{align}

\subsection{PPN orbital radius}
We suppose again that three masses in circular motion 
are in a triangular configuration with a constant angular velocity $\omega$. 
By noting that a vector in the orbital plane 
can be expressed as a linear combination of $\bm{x}_1$ and $\bm{v}_1$, 
Eq.(\ref{PPN-EOM}) becomes 
\begin{align}
- \omega^2 \bm{x}_1 =& - (\omega_N)^2 \bm{x}_1 
+ g_{1} (\omega_N)^2\bm{x}_1 \nonumber\\ 
                                     &+ \frac{\sqrt{3}M}{16 a} \frac{\nu_2 \nu_3 (\nu_2 - \nu_3) (16 \beta -1 -9 \nu_1)}{\nu_2^2 + \nu_2 \nu_3 + \nu_3^2} 
                                      \omega_N\bm{v}_1 ,
                                     \label{EOM-PPN-triangular-1}
\end{align}
where 
Eq. (\ref{omega2}) is used and 
\begin{align}
g_{1} 
= &
\frac{M}{a} 
\left[
\left( 2\beta + \gamma + (\nu_2+\nu_3)(\nu_2+\nu_3-1) 
-\frac{7}{16}\nu_2\nu_3\right)\right.
 \notag\\
&~~~~~~~~
\left.
+\frac{3}{16}\frac{\nu_2\nu_3 \{9\nu_2\nu_3+2(\nu_2+\nu_3)(8\beta-5)\}}{\nu_2^2 + \nu_2\nu_3 + \nu_3^2}
\right] .
\label{g}
\end{align}
By a cyclic permutation, we obtain the similar equations for $M_2$ and $M_3$. 

The second and third terms in the right-had side of Eq. (\ref{EOM-PPN-triangular-1}) 
are the PPN forces. 
The second term is parallel to $\bm{x}_1$, 
whereas the third term is parallel to $\bm{v}_1$ 
Note that 
$\bm{v}_1$ is not parallel to $\bm{x}_1$ 
in circular motion.

The location of the COM in the fully conservative theories of PPN 
\cite{Baker1978,Baker1979}
remains the same as that in the PN approximation of general relativity  
\cite{MTW,LL} 
\begin{eqnarray}
\boldsymbol{G}_{PN} 
= 
\frac{1}{E} \sum\limits_AM_A\boldsymbol{x}_A 
\left[1 + \frac12
\left(v_A^2 - \sum\limits_{B \neq A}\frac{M_B}{R_{AB}}
\right) \right] , 
\label{PN-COM}
\end{eqnarray}
where $E$ is defined as 
\begin{eqnarray}
E \equiv \sum\limits_AM_A\left[1 + \frac12
\left(v_A^2 - \sum\limits_{B \neq A}\frac{M_B}{R_{AB}}
\right) \right] . 
\end{eqnarray}

This coincidence allows us to obtain the PPN orbital radius $\ell^{PPN}_A$ 
around the COM by straightforward calculations. 
The orbital radius of $M_1$ is formally obtained as 
\begin{eqnarray}
(\ell^{PPN}_1)^2 &=& (\ell_1)^2 
 \nonumber \\
&& + \frac{aM}{2}  \left(1 - \frac{a^3 \omega_N^2}{M} \right)
\nonumber\\
&&~~
\times ( - 2\nu_1^2\nu_2^2  - 2\nu_2^2\nu_3^2- 2\nu_3^2\nu_1^2 
\nonumber\\
&&
~~~~~~~
+ 2\nu_1\nu_2^3 + \nu_2\nu_3^3 + \nu_2^3\nu_3 + 2\nu_3^3\nu_1 
 \nonumber \\
&& 
~~~~~~~
 - 2\nu_1^2\nu_2\nu_3 + \nu_1\nu_2^2\nu_3 + \nu_1\nu_2\nu_3^2 )  , 
\label{PPN-ell1}
\end{eqnarray}
and the similar expressions of $\ell^{PPN}_2$ and $\ell^{PPN}_3$ 
for the orbital radius of $M_2$ and $M_3$ are obtained.

Unless the second term of the right-hand side in Eq. (\ref{PPN-ell1}) vanishes, 
the difference between $\ell^{PPN}_1$ and $\ell_1$ would make our computations 
rather complicated.  
However, it 
vanishes 
because $\omega_N$ satisfies Eq. (\ref{omega2}). 
As a result, 
the PPN orbital radius remains the same 
as the Newtonian one. 
Namely, $\ell^{PPN}_{A} = \ell_A$.

\subsection{Equilateral condition}
First, we discuss a condition for an equilateral configuration. 

For Eq. (\ref{EOM-PPN-triangular-1}) to hold, 
the coefficient of the velocity vector $\bm{v_1}$ must vanish, 
because there are no other terms including $\bm{v_1}$. 
The coefficient is proportional to $\nu_2 \nu_3 (\nu_2 - \nu_3)$. 
The same thing is true also of $M_2$ and $M_3$. 
For any value of $\beta$, 
therefore, 
the equilateral configuration in the PPN gravity 
can be present if and only if 
three finite masses are equal or two test masses orbit around one finite mass. 

Note that one can find a very particular value of $\beta$ 
satisfying 
\begin{align}
16\beta -1 - 9\nu_1 = 0 , 
\end{align}
which leads to the vanishing coefficient of the velocity vector $\bm{v_1}$. 
However, this choice is very unlikely, 
because 
the particular value of $\beta$ is dependent on 
the mass ratio $\nu_1$ 
and it is not universal. 
Hence, this case will be ignored.

\subsection{PPN triangular configuration for general masses}
\begin{figure}
\includegraphics[width=7.5cm]{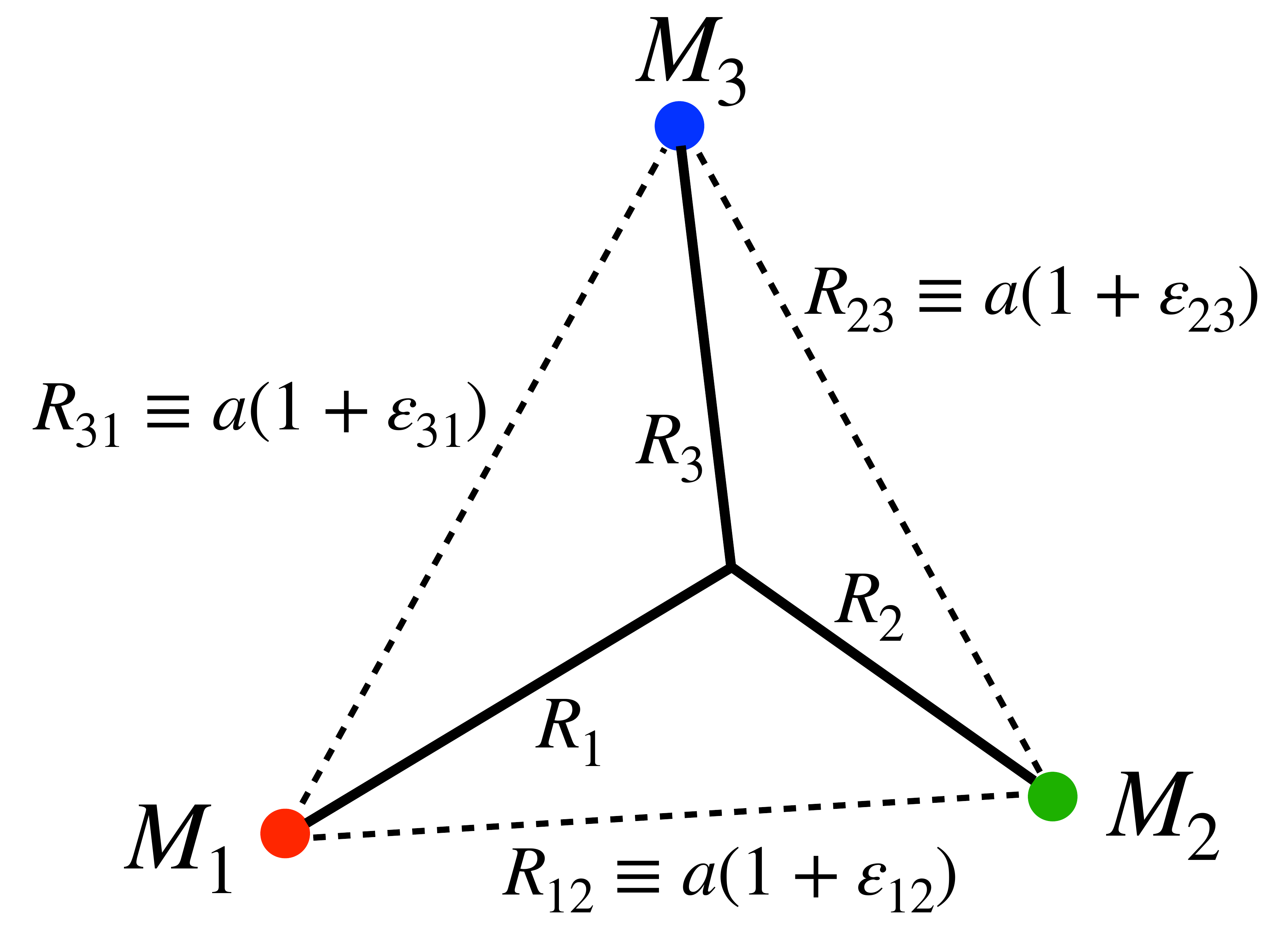}
\caption{
Schematic figure for the PPN triangular configuration of three masses. 
An inequilateral triangle is described by 
the parameter $\varepsilon_{AB}$. 
$R_A$ coincides with $\ell_A$ in the Newtonian limit, 
for which $\varepsilon_{AB}$ vanishes. 
}
\label{figure-triangular}
\end{figure}

Next, let us consider a PPN triangle configuration for general masses. 
For this purpose, 
we introduce a nondimensional parameter 
$\varepsilon_{AB}$ at the PPN order, 
such that each side length of the PPN triangle can be expressed as 
\begin{align} 
R_{AB} = a (1 + \varepsilon_{AB}) . 
\end{align}
The equilateral case is achieved by assuming 
$\varepsilon_{AB} = 0$ for every masses. 
See Figure \ref{figure-triangular} for the PPN triangular configuration.

In order to fix the degree of freedom corresponding to 
a scale transformation, 
we follow Reference \cite{Yamada2012} 
to suppose that the arithmetic mean of the three side lengths 
is unchanged as 
\begin{align}
\frac{R_{12} + R_{23} + R_{31}}{3} = a\bigg[ 1 + \frac{1}{3} (\varepsilon_{12} + \varepsilon_{23} + \varepsilon_{31}) \bigg] . 
\label{mean}
\end{align}
The left-hand side of Eq. (\ref{mean}) is $a$ in the Newtonian case, 
which leads to 
\begin{align}
\varepsilon_{12} + \varepsilon_{23} + \varepsilon_{31} = 0 .
\end{align} 
This is a gauge fixing in $\varepsilon_{AB}$. 

In terms of $\varepsilon_{AB}$, 
Eq. (\ref{EOM-PPN-triangular-1}) is 
rearranged as 
\begin{align}
- \omega^2 \bm{x}_1 
=& 
- (\omega_N)^2 \bm{x}_1 
\notag\\                                         
&  
 - \frac{3}{2} \frac{(\omega_N)^2}{\nu_2^2 + \nu_2 \nu_3 + \nu_3^2} 
 \notag\\
& 
\times\bigg[ \bigg\{
                                         \nu_2 (\nu_1 - \nu_2 -1)\varepsilon_{12} + \nu_3 (\nu_1 -\nu_3 -1) \varepsilon_{31} \bigg\} \bm{x}_1 \nonumber\\
                                         &~~~~~~
                                          + \sqrt{3} \nu_2 \nu_3 (\varepsilon_{12} - \varepsilon_{31}) \frac{\bm{v}_1}{\omega_N}
                                         \bigg] + \bm{\delta}_1 ,
                                         \label{EOM-PPN-triangular-2}
\end{align}
where 
\begin{align}
\bm{\delta}_1 =
&  g_{1} (\omega_N)^2\bm{x}_1 
+
\frac{\sqrt{3}M\nu_2 \nu_3 (\nu_2 - \nu_3)(16 \beta -1 -9 \nu_1)}{16 a (\nu_2^2 + \nu_2 \nu_3 + \nu_3^2)} 
\omega_N\bm{v}_1  .
\end{align}
By a cyclic permutation, 
the equations for $M_2$ and $M_3$ can be obtained.

A triangular equilibrium configuration can exist 
if and only if 
the two conditions (A) and (B) are simultaneously satisfied; 
(A) Each mass satisfies Eq. (\ref{EOM-PPN-triangular-2}), 
and 
(B) the configuration is unchanged in time.

Eq. (\ref{EOM-PPN-triangular-2}) is the equation of motion for $M_1$. 
To be more accurate, therefore, 
$\omega$ in Eq. (\ref{EOM-PPN-triangular-2}) should be denoted as $\omega_1$. 
Similarly, we introduce $\omega_2$ and $\omega_3$ 
in the equations of motion for $M_2$ and $M_3$, respectively. 
Then, Condition (B) 
means $\omega_1 = \omega_2 = \omega_3$. 

Condition (A) is equivalent to Condition (A2); 
The coefficient of $\bm{v}_A$ in the equation of motion 
vanishes as 
 \begin{align}
\varepsilon_{12} - \varepsilon_{31} - \frac{M}{24 a} (\nu_2 - \nu_3) (16  \beta-1      -9\nu_1 ) = 0 , 
\label{1}
\end{align}
\begin{align}
\varepsilon_{23} - \varepsilon_{21} - \frac{M}{24 a} (\nu_3 - \nu_1) (16  \beta   -1   -9 \nu_2 ) = 0 ,
\label{2}
\end{align}
\begin{align}
\varepsilon_{31} - \varepsilon_{23} - \frac{M}{24 a} (\nu_1 - \nu_2) (16  \beta -1     -9\nu_3 ) = 0 .
\label{3}
\end{align}

From Eqs. (\ref{1})-(\ref{3}) and the gauge fixing as 
$\varepsilon_{12} + \varepsilon_{23} + \varepsilon_{31} = 0$, 
we obtain 
\begin{align}
\varepsilon_{12} 
= \frac{M}{72a} 
&\bigg[
(\nu_2 - \nu_3) (16 \beta  -1 -9 \nu_1) 
\notag\\
& - (\nu_3 - \nu_1) (16 \beta -1  -9 \nu_2)
\bigg] ,
\label{varepsilon12}
\end{align}
\begin{align}
\varepsilon_{23} = \frac{M}{72a} 
&\bigg[
(\nu_3 - \nu_1) (16 \beta  -1 -9 \nu_2) 
\notag\\
&- (\nu_1 - \nu_2) (16 \beta  -1 -9 \nu_3)
\bigg] ,
\label{varepsilon23}
\end{align}
and 
\begin{align}
\varepsilon_{31} = \frac{M}{72a} 
&\bigg[
(\nu_1 - \nu_2) (16 \beta -1  -9 \nu_3) 
\notag\\
&- (\nu_2 - \nu_3) (16 \beta  -1 -9 \nu_1)
\bigg] . 
\label{varepsilon31}
\end{align}

Therefore, the PPN triangle is inequilateral  depending on $\beta$ via $\varepsilon_{AB}$ 
but not on $\gamma$. 
This suggests that also the PPN Lagrange points 
corresponding to $L_4$ and $L_5$ 
are sensitive to $\beta$ but are free from $\gamma$, 
as shown in Section IV.

It follows that Eqs. (\ref{varepsilon12})-(\ref{varepsilon31}) recover 
 the PN counterpart of  
 Eq. (26)-(28) of Reference \cite{Yamada2012} 
if and only if $\beta = 1$. 
The uniqueness is because the PPN parameter is only $\beta$ for 
three equations as Eqs. (\ref{varepsilon12})-(\ref{varepsilon31}).

Condition (B) is satisfied, 
if $\omega_1 = \omega_2 = \omega_3 \equiv \omega_{PPN}$, 
where $\omega_{PPN}$ means the angular velocity of the PPN configuration.  
By substituting Eqs. (\ref{varepsilon12}) and (\ref{varepsilon31}) 
into Eq. (\ref{EOM-PPN-triangular-2}), 
$\omega_{PPN}$ is obtained as 
\begin{align}
\omega_{PPN} = \omega_N \left(1 + \delta_{\omega} \right) ,
\end{align}
where, 
by using Eq. (\ref{g}), 
the PPN correction $\delta_{\omega}$ is 
\begin{align}
\delta_{\omega} 
=& 
\frac34 \frac{\nu_2 (\nu_1 - \nu_2 -1)\varepsilon_{12} + \nu_3 (\nu_1 -\nu_3 -1) \varepsilon_{31}}{\nu_2^2 + \nu_2 \nu_3 + \nu_3^2} 
- \frac12 g_{1} 
\notag\\
=& 
- \frac{M}{48a} \{64\beta +24\gamma -1 -42 (\nu_1 \nu_2 + \nu_2 \nu_3 + \nu_3 \nu_1)  \} .
\label{deltaomega}
\end{align}

There is a symmetry among $M_1, M_2, M_3$ in the second line of Eq. (\ref{deltaomega}), 
which means that $\delta_{\omega}$ is the same for all bodies. 
Condition (B) is thus satisfied.

%%%%
\section{PPN corrections to the Lagrange points}
\subsection{PPN Lagrange points $L_1$, $L_2$ and $L_3$}
In this section, we discuss PPN modifications of the Lagrange points 
that are originally defined in the restricted three-body problem in Newton gravity. 
We choose $\nu_A = 1-\nu$, $\nu_B = \nu$ and $\nu_C=0$, 
where $\nu$ is the the mass ratio of the secondary object (a planet). 

First, we seek PPN corrections to $L_1, L_2$ and $L_3$. 
There are three choices of how to 
correspond 
$M_1, M_2$ and $M_3$ 
to the Sun, a planet and a test mass 
in the collinear configuration. 
Indeed the three choices lead to the Lagrange points $L_1$, $L_2$ and $L_3$. 

We consider the collinear solution by Eq. (\ref{7theq}). 
We denote the physical root for Eq. (\ref{7theq})  
as $z = z_N (1 + \varepsilon)$ for the Newtonian root $z_N$ 
with using a small parameter $\varepsilon$ ($|\varepsilon| \ll 1$) 
at the PPN order. 
We substitute $z$ into Eq. (\ref{7theq}) and rearrange it 
to obtain $\varepsilon$ as 
\begin{align}
\varepsilon 
= -
\cfrac{\sum\limits_{k=0}^7 A^{PPN}_{k} (z_N)^k}{\sum\limits_{k=1}^6 k A^N_{k} (z_N)^k}, 
\label{varepsilon}
\end{align}
where 
$O(\varepsilon^2)$ is discarded because of being at the 2PN order, and 
$A^N_{k}$ and $A^{PPN}_{k}$ denote the Newtonian and PPN parts of $A_k$, 
respectively, as $A_k = A^N_{k} + \varepsilon A^{PPN}_{k}$ 
($A^N_{0} = 0$ and $A^N_{7}=0$ because there are no counterparts 
in the Newtonian case).

Eq. (\ref{varepsilon}) is used for 
calculating the PPN corrections to $L_1, L_2$ and $L_3$. 
The PPN displacement from 
the Newtonian Lagrange point 
$L_1$ 
is thus obtained as 
\begin{align}
\delta_{PPN} R_{23} 
&\equiv R_{23} - (R_{23})_N 
\notag\\
&= \frac{\varepsilon z_N}{(1 + z_N)^2}\ell + O(\ell\varepsilon^2) ,
\end{align}
where $M_1$, $M_2$ and $M_3$ are chosen 
as a planet, a test mass and the Sun, respectively. 

Similarly, 
the PPN displacement from 
the Newtonian Lagrange point
$L_2$  
becomes 
\begin{align}
\delta_{PPN} R_{31} 
&\equiv R_{31} - (R_{31})_N 
\notag\\
&= \frac{\varepsilon z_N}{(1 + z_N)}  \ell + O(\ell\varepsilon^2) ,
\end{align}
where 
$M_1$, $M_2$ and $M_3$ are chosen 
as the Sun, a planet and a test mass, respectively. 
The PPN displacement from 
the Newtonian Lagrange point 
$L_3$ 
is 
\begin{align}
\delta_{PPN} R_{23} 
&\equiv R_{23} - (R_{23})_N 
\notag\\
&= \frac{\varepsilon z_N}{(1 + z_N)}  \ell + O(\ell\varepsilon^2) ,
\end{align}
where 
$M_1$, $M_2$ and $M_3$ are chosen 
as a planet, the Sun and a test mass, respectively. 
Here, a value of $z_N$ depends on $L_1$, $L_2$ or $L_3$, 
which is given by Eq. (\ref{5th}).

\subsection{PPN Lagrange points $L_4$ and $L_5$}
Next, we discuss PPN corrections to 
the  Lagrange points $L_4$ and $L_5$, 
for which we consider the PPN triangular solution. 
Let $a$ denote 
the orbital separation between the primary object and the secondary one, 
which equals to $R_{12} = \ell(1 + \varepsilon_{12})$. 
Therefore, $\ell = a (1 - \varepsilon_{12}) + O(a \varepsilon^2)$, 
where $\varepsilon^2$ denotes the second order in $\varepsilon_{AB}$. 
By using this for $R_{23}$ and $R_{31}$, 
we obtain
$R_{23} = a (1 + \varepsilon_{23} - \varepsilon_{12}) + O(a \varepsilon^2)$, 
and 
$R_{31} = a (1 + \varepsilon_{31} - \varepsilon_{12}) + O(a \varepsilon^2)$. 

The PPN displacement from
the Newtonian Lagrange point 
$L_4$ (and $L_5$) 
with respect to the Sun  
is obtained as 
\begin{align}
\delta_{PPN} R_{31} 
\equiv& R_{31} - a 
\notag\\
=& 
a (\varepsilon_{31}-\varepsilon_{12}) + O(a\varepsilon^2) 
\notag\\
=& 
- \frac{\nu(16\beta - 10 + 9\nu)}{24}  M
\notag\\
&+ O\left(\frac{M^2}{a}\right) , 
\label{deltaL4}
\end{align}
where 
$\nu_1 = 1-\nu$, $\nu_2 = \nu$ and $\nu_3=0$ are used 
in the last line. 

In the similar manner, 
the PPN displacement from 
the Newtonian Lagrange point 
$L_4$ (and $L_5$) 
with respect to the planet 
\begin{align}
\delta_{PPN} R_{23} 
&\equiv R_{23} - a 
\notag\\
&= a (\varepsilon_{23}-\varepsilon_{12}) + O(a\varepsilon^2) 
\notag\\
&= - \frac{(1-\nu) (16\beta -1 -9\nu)}{24} M + O\left(\frac{M^2}{a}\right) . 
\label{deltaL5}
\end{align}

Eq. (\ref{deltaL5}) can be obtained more easily from Eq. (\ref{deltaL4}) 
if the correspondence as $1-\nu \leftrightarrow \nu$ is used.

\begin{table}[h]
\caption{The PPN displacement from 
the Newtonian Lagrange points 
of the Sun-Jupiter system.
The PPN corrections to $L_1$, $L_2$, $L_3$ and $L_4$ are listed in this table, 
where the sign convention for $L_1$, $L_2$, $L_3$ 
is chosen along the direction from the Sun to the Jupiter, 
and the correction to $L_5$ is identical to that to $L_4$. 
The PPN displacement for $L_4$ 
is two-dimensional and 
hence they are indicated by the deviations from the Sun 
and from the Jupiter. 
}
  \begin{center}
    \begin{tabular}{l|ll}
\hline
Lagrange points\: &  & PPN displacement [m] \\
\hline 
$L_1$& &$-0.000051 + 40.00 \beta - 9.905\gamma$  \\
$L_2$& &$0.000040 - 50.27 \beta + 12.40 \gamma$  \\
$L_3$& &$0.000122 + 1.424 \beta + 0.01882 \gamma$  \\
$L_4(L_5)$-Sun& &$-0.05875 \times(-9.991 + 16 \beta)$  \\
$L_4(L_5)$-Jupiter & &$-61.53 \times(-1 + 16 \beta)$  \\
\hline
    \end{tabular}
  \end{center}
\label{Lagrange}
\end{table}

\subsection{Example: the Sun-Jupiter case}
The PPN corrections to the $L_1$, $L_2$ and $L_3$ 
can be expressed as a linear function in $\beta$ and $\gamma$. 
The PPN corrections to $L_4$ and $L_5$ are 
in a linear function only of $\beta$. 
The results for the Sun-Jupiter system 
are summarized in Table \ref{Lagrange}, 
where the sign convention is chosen along the direction from the Sun to a planet.

Before closing this section, we mention gravitational experiments. 
The lunar laser ranging experiment put a constraint 
on $\eta \equiv 4\beta - \gamma -3$ 
as $|\eta| < O(10^{-4})$ 
\cite{Williams1996,Williams2004}. 
If one wish to constrain $1 - \beta$ at the level of $O(10^{-4})$ 
by using the location of the Lagrange points, 
the Lagrange point accuracy of about a few millimeters 
(e.g. for $L_4$) 
is needed in the solar system, 
though this is very unlikely in the near future. 

On the other hand, possible PPN corrections in a three-body system 
may be relevant with relativistic astrophysics 
in e.g. 
a relativistic hierarchical triple system 
and a supermassive black hole with a compact binary 
\cite{Rosswog,Suzuki,Fang,Kunz2021,Kunz2022}.  
This subject is beyond the scope of the present paper.

%%%%%
\section{Conclusion}
The coplanar 
and circular  
three-body problem was investigated 
for a class of fully conservative theories in the PPN formalism, 
characterized by the Eddington-Robertson parameters $\beta$ and $\gamma$.  

The collinear configuration 
can exist for arbitrary mass ratio, $\beta$ and $\gamma$. 
On the other hand, 
the PPN triangular configuration 
depends on the nonlinearity parameter $\beta$ 
but not on $\gamma$. 
This is far from trivial, because 
the parameter $\beta$ is not separable from $\gamma$ 
apparently at the level of Eq. (\ref{PPN-EOM}). 
For any value of $\beta$, 
the equilateral configuration in the PPN gravity is possible,  
if and only if 
three finite masses are equal or two test masses orbit around one finite mass. 
For general mass cases, 
the PPN triangle is not equilateral. 

We showed also that 
the PPN displacements from 
the Newtonian Lagrange points 
$L_1$, $L_2$ and $L_3$  
depend on both $\beta$ and $\gamma$, 
while those to $L_4$ and $L_5$ 
rely only upon $\beta$. 
It is left for future to study the stability of the PPN configurations.

\section{Acknowledgments}
We thank Kei Yamada and Yuuiti Sendouda 
for fruitful conversations. 
This work was supported 
in part by Japan Science and Technology Agency (JST) SPRING, 
Grant Number, JPMJSP2152 (Y.N.), 
and 
in part by Japan Society for the Promotion of Science (JSPS) 
Grant-in-Aid for Scientific Research, 
No. 20K03963 (H.A.).

\end{document}